\documentclass[twocolumn,showpacs,amsmath,amssymb,aps,prc]{revtex4}

\usepackage[dvips]{graphicx}
\usepackage{dcolumn}
\usepackage{bm}
\begin{document}

\newcommand{\Lu}{$^{176}{\rm Lu}$}
\newcommand{\Ne}{$^{22}{\rm Ne}$}
\newcommand{\Hf}{$^{176}{\rm Hf}$}
\newcommand{\ex}{E$_{\rm x}$}
\newcommand{\jpi}{$J^{\pi}$}
\newcommand{\tnine}{T$_{9}$}
\newcommand{\sproc}{\textit{s}-process}
\preprint{APS/123-QED}
\title{Thermal Equilibration of $^{176}$Lu via $K$-Mixing}
\author{Vadas~Gintautas$^{1}$\footnote{present address: The University of Illinois, Urbana, IL 61801-3080, USA}, Arthur~E.~Champagne$^{1,2}$,  Filip~G. Kondev$^{3}$ and Richard Longland$^{1,2}$}
\affiliation{$^{1}$The  University of North Carolina at Chapel Hill, Chapel 
Hill, North Carolina 27599-3255, USA\\
$^{2}$Triangle Universities Nuclear Laboratory, Durham, North Carolina, 
27708-0308, USA\\
$^{3}$Argonne National Laboratory, Argonne, Il 60439, USA}

\date{\today}
\begin{abstract}
In astrophysical environments, the long-lived ($T_{1/2}$ = 37.6 Gy) ground state of \Lu\ can communicate with a short-lived ($T_{1/2}$ = 3.664 h) isomeric level through thermal excitations.  Thus, the lifetime of \Lu\ in an astrophysical environment can be quite different than in the laboratory. We examine the possibility that the rate of equilibration can be enhanced via $K$-mixing of two levels near \ex\ = 725 keV and estimate the relevant $\gamma$-decay rates.  We use this result to illustrate the effect of $K$-mixing on the effective stellar half-life. We also present a network calculation that includes the equilibrating transitions allowed by $K$-mixing. Even a small  amount of $K$-mixing will ensure that \Lu\  reaches at least a quasi-equilibrium during an \sproc\ triggered by the \Ne\ neutron source.
\end{abstract}
\maketitle
\section{Introduction}
The nucleus \Lu\ is produced almost exclusively by slow neutron capture in the \textit{s}-process. It was originally thought that by virtue of its long half-life ($T_{1/2}$ = 37.6 Gy), \Lu\ could serve as a chronometer of galactic nucleosynthesis. However, it has been shown~\citep{klay91b,lesko91} that the ground state of \Lu\ can reach at least a quasi-equilibrium with a short-lived isomeric state at \ex\ = 122.9 keV ($T_{1/2}$ = 3.664 h; all excitation energies and other nuclear structure parameters are taken from the {\sc ENDSF} compilation~\cite{endsf}) through a mediating level at \ex\ = 838.6 keV. As a result, the astrophysical half-life of \Lu\  is strongly dependent on temperature. This precludes the use of \Lu\  as a \textit{chronometer}, but does imply that it may serve instead as an \textit{s}-process \textit{thermometer}.

\begin{figure}[htb]
	\centering
		\includegraphics[width=0.5\textwidth]{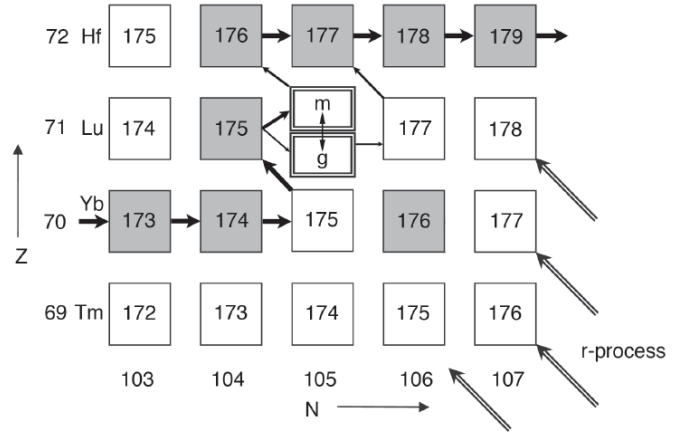}
	\caption{The \textit{s}-process path in the vicinity of \Lu. Stable (or extremely llong-lived) nuclei are shown in grey. The ground state of \Lu\ and the isomeric state are labeled as g and m, respectively.}
	\label{fig:sprocess}
\end{figure}

The \sproc\ in the in $A$ = 176 mass region is shown in Figure \ref{fig:sprocess}. Production of the isomer, either by neutron capture or by thermal excitation of the ground state leads directly (via $\beta$-decay) to \Hf. In contrast, the ground state, which is also produced directly by neutron capture or by thermal excitation of the isomer,  will undergo neutron capture to $^{177}{\rm Lu}$, which then $\beta$-decays to  $^{177}{\rm Hf}$. This temperature-dependent branching in the reaction flow should be reflected in the abundances in this mass region. In fact, consideration of isotopic abundances strongly suggests that the isomer and ground state must communicate during the \textit{s}-process: As discussed by Klay \textit{et al.}~\citep{klay91b} (and references therein),
measurements of the partial neutron capture cross section producing the isomer show that it is too large to be consistent with the observed abundance of \Lu\ in the ground state.  Recent measurements reinforce this conclusion~\cite{wiss}. Studies of the branching of the \textit{s}-process at $A$ = 176 also indicate that there is more \textit{s}-process flow through the ground state than predicted by the production rate of the ground state~\cite{beer84} via neutron capture in the classical picture.  Taken together, these observations suggest the presence of an additional creation mechanism for the ground state. Neutron capture produces an isomer to ground state ratio of approximately 5.7~\cite{kad}. At typical \textit{s}-process temperatures (about 1--3 $\times$ 10$^{8}$ K, or \tnine\ = 0.1--0.3), the thermal equilibrium ratio is 1.3 $\times$ 10$^{-7}$--1.7 $\times$ 10$^{-3}$. In other words, if thermal equilibrium pertains, then most of the \sproc\ flow would ultimately pass through the ground state.

Measurements of $\gamma$-decays by Klay \textit{et al.}~\citep{klay91a}, Lesko  \textit{et al.}~\citep{lesko91} and Vanhorenbeeck \textit{et al.}~\citep{van} have uncovered several paths linking the ground state and the isomer, which would be active during an \sproc\ triggered by the \Ne\ neutron source. In this paper we will examine the possibility that $K$-mixing provides a more efficient way for the ground state and isomer to reach equilibrium. This mechanism is not limited to the case of \Lu\ and should operate in other \sproc\ nuclei.

\section{Equilibration of \Lu}
Direct $\gamma$-ray decay of the isomer (\jpi\ = 1$^{-}$, $K^{\pi}$ = 0$^{-}$) to the ground state (\jpi\ = 7$^{-}$, $K^{\pi}$ = 7$^{-}$) is strongly suppressed by their large difference in angular momentum. Also, because of the approximate conservation of the $K$ quantum number,  $\gamma$-decay is further hindered by a factor of 100 per degree of $K$-forbiddenness~\cite{lobner68} (i.e., by a factor 10$^{2\nu}$, where $\nu$ $\approx$ $|\Delta K-\lambda|$ and $\lambda$ is the $\gamma$-ray multipolarity). In general, the approach to equilibrium under stellar conditions can be inferred by comparing the time scales for $\gamma$-ray absorption and emission to the time scales for nuclear reactions and decays. States that are strongly linked by downward (and therefore upward) transitions, such as members of a given rotational band, will quickly reach full thermal equilibrium and will evolve together. States that are more weakly connected may not reach thermal equilibrium if other nuclear processes are faster. In the case of \Lu, the members of the ground-state band ($\pi 7/2\left[404\right] + \nu 7/2\left[514\right], K^{\pi}$ = 7$^{-}$) will quickly reach thermal equilibrium. For example, at \tnine\ = 0.3, the 184.1-keV, \jpi\ = 8$^{-}$ state will equilibrate with the ground state with a time constant of 21 ns. Side bands that feed the ground-state band will also equilibrate with the ground-state band. Similarly, the states in the isomer band ($\pi 7/2\left[404\right] - \nu 7/2\left[514\right], K^{\pi}$ = 0$^{-}$) and associated side bands will rapidly form an equilibrium cluster. At this point \Lu\ can be thought of as a quasi-equilibrium ensemble consisting of two separate, equilibrated clusters (the ground-state and isomer clusters). New groups of states in local equilibrium will form, presumably via $K$-allowed transitions (either upward or downward) from states in the original equilibrium clusters to those in other bands. If the time scale for linking transitions is short (compared to the time scale for neutron capture or $\beta$-decay), then these new states will  reach equilibrium with the original cluster. Otherwise, they will form separate groups in local equilibrium. Eventually, the clusters originating from the ground state will reach those based on the isomer and produce quasi-equilibrium between the ground state and isomer. This process will be discussed in more detail in Sec. V.

The previous $\gamma$-decay studies of \Lu,~\citep{klay91a,lesko91} clearly show that the \jpi\ = 5$^{-}$ state at \ex\ = 838.6 keV ($K$ = 4) decays directly to the ground state via a strong  E2 transition \textit{and} via cascades that ultimately end in the isomer (more recent measurements~\cite{van} show evidence for mediating states at somewhat higher excitation energy). In a stellar plasma, this state would be populated from the isomer in the manner described above, via 3--4 interband transitions, and primarily via a single photon from the ground state. Interestingly, several of the strongest transitions from this level (including the ground-state branch) are $K$-forbidden~\cite{doll}, which implies that the 838.6-keV state may be mixed with other nearby 5$^{-}$ states. 

\begin{figure*}[htb]
	\centering
		\includegraphics[width=.75\textwidth]{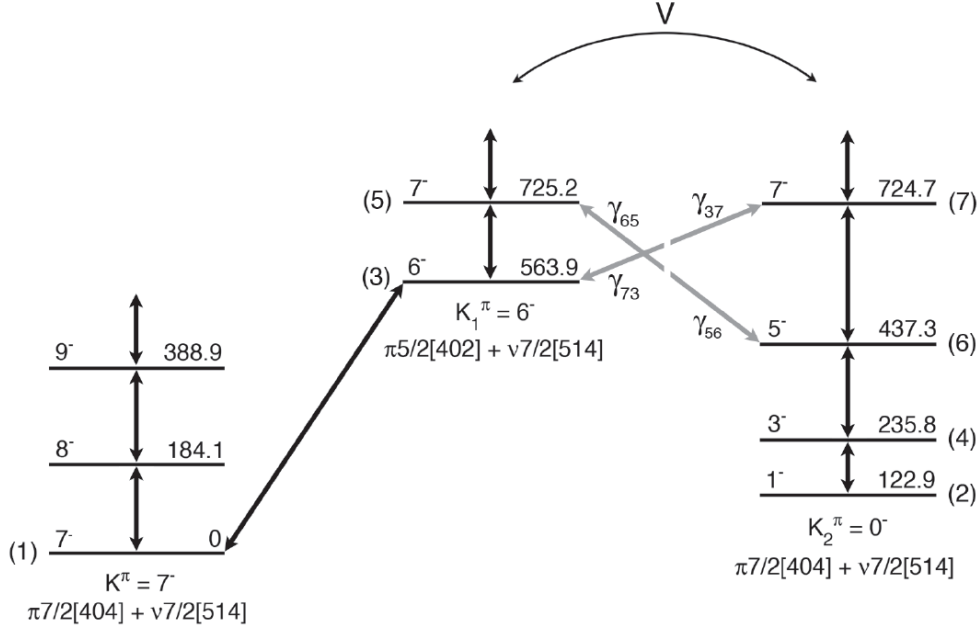}
	\caption{Proposed equilibration scheme. States are listed according to their \jpi\ and excitation energies (in keV). The numbers in parentheses denote the states of interest. Downward $\gamma$-transitions have been observed for the links represented by black arrows. The matrix element that describes the mixing of the 7$^{-}$ states [labeled (5) and (7)] in the $K$$^{\pi}$ = 6${-}$ and $K$$^{\pi}$ = 0${-}$ bands is denoted by $V$. The 2 bands are potentially linked by $\gamma$-transitions between states (5) and (6) and between (3) and (7) (shown in grey).}
	\label{fig:equilstates}
\end{figure*}

Klay \textit{et al.}~\citep{klay91b} discuss the possibility that weak, undetected $K$-forbidden transitions may also provide an equilibration path, but argue that these contributions are small compared to the observed route. However, the effect of any such transition could be enhanced by $K$-mixing, which would largely circumvent the $K$-selection rule. As mentioned above, this mechanism appears to be at work for equilibration through the 838.6-keV state and there is clear evidence for $K$-mixing at higher excitation energies \cite{mcg}. Although the lower-energy members of the isomer band have ratios of dynamic to kinematic moments of inertia that are nearly unity (indicating little rotational perturbation), chance degeneracies can also lead to $K$-mixing and thus it seems at least plausible that the 7$^{-}$ states at \ex\ = 724.7 keV  ($\pi 7/2\left[404\right] - \nu 7/2\left[514\right], $K$^{\pi}$ = 0$^{-}$) and 725.2 keV  ($\pi 5/2\left[402\right] - \nu 7/2\left[514\right], K^{\pi}$ = 6$^{-}$) will mix. The former state is a member of the isomer band whereas the latter state is a member of a band that can be populated by thermal excitation of the ground state. If these states do indeed mix, then the ground state and isomer are linked by a series of $K$-allowed transitions. As pointed out by Kondev  {\em et al.}~\cite{kon} and McGoram {\em et al.}~\cite{mcg}, even small amounts of mixing can significantly increase the rate of transitions that would otherwise be $K$-hindered. These transitions could be much weaker than the unmixed transitions and thus would have escaped notice in previous decay studies. However, since the mediating states would be at lower excitation energy, $\approx$725 keV rather than at 839 keV, they might be produced more efficiently from the stellar photon bath. This scenario for equilibration is illustrated in Figure~\ref{fig:equilstates}.

\section{$K$-mixing of 2 levels and $\gamma$-decay}

If we assume that 2 states of a given \jpi, but different $K$ ($K_{1}$ and $K_{2}$) are mixed, then the resultant wavefunction for each state can be written schematically as
\begin{eqnarray}
\left| J^{\pi},\bar{K}_{1}\right> = \alpha \left| J^{\pi},K_{1}\right> - \beta  \left| J^{\pi},K_{2}\right>  
\end{eqnarray}
and
\begin{eqnarray}
\left| J^{\pi},\bar{K}_{2}\right> = \alpha \left| J^{\pi},K_{1}\right> + \beta  \left| J^{\pi},K_{2}\right>,
\end{eqnarray}
where
\begin{equation}
\alpha^{2} + \beta^{2} = 1.
\end{equation}
In our case, \jpi\ = 7$^{-}$ ; $\bar{K}_{1}$ and $\bar{K}_{2}$ refer to the mixed-$K$ nature of the states (where $K$ is no longer a good quantum number); the pure $K$ states have $K_{1}$ = 6, and $K_{2}$ = 0. 

The matrix element for the E2 transition linking e.g., states (5) and (6) in Figure~\ref{fig:equilstates} [henceforth, transition a$\rightarrow$b will be denoted as (ab)] is 
\begin{align}
\left<  5^{-}, 0 \left\| M(E2) \right\| 7^{-}, \bar{K}_{1} \right> &= \alpha \left< 5^{-}, 0 \left\| M(E2) \right\| 7^{-}, 6 \right>\nonumber \\
&-\beta \left< 5^{-}, 0 \left\| M(E2) \right\| 7^{-}, 0 \right> 
\end{align}
where M(E2) is the electric quadrupole operator. The first term represents a $K$-forbidden decay, which can be ignored. Assuming that T$^{\gamma}_{1/2}$/T$^{W}_{1/2}$ = 100$^{\nu}$ (where T$^{W}_{1/2}$ is the Weisskopf estimate for the partial half-life and $\nu$ is the $K$-hindrance factor introduced above), T$^{\gamma}_{1/2}$(56) = 488 ms, which is much longer than for the allowed decays calculated below. Since
\begin{align}
&B(E2;J,K \rightarrow J-2,K) = \nonumber\\
&\qquad \frac{1}{2J+1} \left|\left<J-2,K\left\| M(E2) \right\| J,K\right>\right|^2,
\end{align}
we have 
\begin{equation}
B(E2;7^{-},\bar{K}_{1} \rightarrow 5^{-},0) =\beta^{2}\,B(E2;7^{-},0 \rightarrow 5^{-},0).
\end{equation}
From the relationship between $B(E2)$ and the partial half-life for the transition, we finally obtain
\begin{equation}
T^{\gamma}_{1/2}(56) = \left(\frac{E_\gamma(76)}{E_\gamma(56)}\right)^{5} \; \frac{T^{\gamma}_{1/2}(76)}{(\beta\\/\alpha)^{2}} \approx \frac{T^{\gamma}_{1/2}(76)}{(\beta\\/\alpha)^{2}}.
\label{eqn:kmix1}
\end{equation}
In other words, the $K$-mixed (56) transition can be calculated from the partial half-life for the observed (76) transition if the mixing amplitude $\beta$ can be determined.

A similar approach can be used for the transition between levels (7) and (3). However one difference is that the (73) and (53) transitions can be of mixed M1 + E2 character. In this case
\begin{align}
\frac{1}{T^{\gamma}_{1/2}(73)} &= \frac{1}{T^{\gamma}_{1/2}(M1;73)} + \frac{1}{T^{\gamma}_{1/2}(E2;73)} \nonumber  \\
&= \frac{1 + \delta^{2}(73)}{T^{\gamma}_{1/2}(M1;73)},
\end{align}
where $\delta(73)$ is the E2/M1 mixing ratio. Assuming that $\delta(73)$ = $\delta(53)$, then
\begin{equation}
T^{\gamma}_{1/2}(73) = \left(\frac{E_\gamma(53)}{E_\gamma(73)}\right)^{5} \; \frac{T^{\gamma}_{1/2}(53)}{(\beta\\/\alpha)^{2}} \approx \frac{T^{\gamma}_{1/2}(53)}{(\beta\\/\alpha)^{2}}.
\label{eqn:kmix2}
\end{equation}

\section{Estimated $\gamma$-decay rates}
The partial half-lives for the observed (53) and (76) transitions have not been measured and so we have calculated them using a semi-empirical approach. For the (76) $E2$ transition, we use the relationship between $B(E2)$ and the intrinsic quadrupole moment $Q_{0}$:
\begin{equation}
B(E2;J,K \rightarrow J-2,K) = \frac{5}{16\pi}\,Q_{0}^{2}\,\left|\left<JK20|J-2K\right>\right|^{2}
\label{eqn:kmix3}
\end{equation}
where $Q_{0}$ = 7.35(4) $e$b is calculated using the measured $Q$ = 4.92(3) $e$b~\cite{endsf} for the ground-state band and the relationship
\begin{equation}
Q = \frac{3K^{2} - J(J + 1)}{(J + 1)(2J + 3)}Q_{0}.
\end{equation}
The remaining term in Eqn.~\ref{eqn:kmix3} is a Clebsch-Gordan coefficient. This yields $B(E2)$ = 1.734 $e$b$^{2}$, which implies  T$^{\gamma}_{1/2}(76)$ = 16.7/$\alpha^{2}$ ps. For the (53) M1 transition, we make use of
\begin{align}
&B(M1;J,K \rightarrow J-1,K) =   \nonumber  \\
&\qquad \frac{3}{4\pi}\,\left(g_{K} - g_{R}\right)^{2}\,K^{2}\,\left|\left<JK0|J-1K\right>\right|^{2}.
\end{align}
The gyromagnetic ratios $g_{K}$($\pi$5/2[402]) = 1.57 and $g_{K}$($\nu$7/2[514]) = 0.30 were calculated using deformed Woods-Saxon wave functions with deformation parameters $\beta_{2}$ = 0.278, 
$\beta_{4}$ = -0.071, and $\beta_{6}$ = -0.009~\cite{moller}. The collective gyromagnetic ratio $g_{R}$ was calculated from
\begin{equation}
\mu = g_{R}J + \left(g_{K} - g_{R}\right)\,\frac{K^{2}}{J + 1}
\end{equation}
with $\mu$ = +3.1692(45) $\mu_{N}$~\cite{endsf} and $\left| \left(g_{K} - g_{R}\right)\right|$ = 0.175(7)~\cite{anu}. The resulting value, $g_{R}$ = 0.30 was assumed to be the same for all bands. From this, we obtain $B(M1)$ = 0.298 $\mu_{N}^{2}$ and T$^{\gamma}_{1/2}(M1)$ = 31.5 ps. However, since this transition can be of mixed M1 + E2 multipolarity, 
\begin{equation}
T^{\gamma}_{1/2} = \frac{T^{\gamma}_{1/2}(M1)}{1 + \delta(53)^{2}},
\end{equation}
where
\begin{equation}
\delta(53)  = 0.933\, \frac{Q_{0}}{ \left(g_{K} - g_{R}\right)}\, \frac{E_{\gamma}(53) (MeV)}{\sqrt{J^{2}-1}} = 0.3.
\end{equation}
Therefore, T$^{\gamma}_{1/2}(53)$ = 28.9/$\alpha^{2}$ ps.

The partial half-lives for the $K$-mixed (56) and (73) transitions can be calculated using Eqns.~\ref{eqn:kmix1} and~\ref{eqn:kmix2}. 
The mixing matrix element $V$ and the amplitudes $\alpha$ and $\beta$ are related by
\begin{equation}
V = \alpha \times \beta \times \Delta E_{x},
\label{eqn:matrix}
\end{equation}
where 
$\Delta E_{x}$ is the energy difference between the two 7$^{-}$ levels (517 eV). Given that $\alpha^{2}$ + $\beta^{2}$ = 1, Eqn.~\ref{eqn:matrix} can be solved for $\beta^{2}$:
\begin{equation}
\beta^{2} = \frac{2\left(\frac{V}{\Delta E_{x}}\right)^{2}}{1 + \left[1-4\left(\frac{V}{\Delta E_{x}}\right)^{2}\right]^{1/2}}.
\end{equation}
Unfortunately, $V$ is unknown and can take values up to $\Delta E_{x}$/2 = 258.5 eV. Consequently, we have calculated partial half-lives and branching ratios for different values of $V$, which are listed in Table~\ref{tab:results}.
\begin{table}[h]
\caption{\label{tab:results}Mixing matrix elements and partial half-lives for $K$-mixed transitions}
\begin{ruledtabular}
\begin{tabular}{cccccc}
 $V$ & $\beta^{2}$ & T$^{\gamma}_{1/2}(76)$  & T$^{\gamma}_{1/2}(56)$  & T$^{\gamma}_{1/2}(53)$ & T$^{\gamma}_{1/2}(73)$ \\ 
(eV) & (\%) & (ps) & (ps) & (ps) & (ps) \\ 
 \hline
 258.5  & 50.0 & 3.34e+1 & 3.34e+1 & 5.78e+01 & 5.78e+01 \\
  150  & 9.28 & 1.84e+1 & 1.80e+02 & 3.19e+01& 3.11e+02 \\
 100  & 3.89 & 1.74e+1 & 4.29e+02 & 3.01e+01& 7.43e+02 \\
 50  & 0.944 & 1.69e+1 & 1.77e+03 & 2.92e+01 & 3.06e+03 \\
 10  & 3.74e-02 & 1.671e+1 & 4.47e+04 & 2.89e+01 & 7.73e+04 \\
 5  & 9.40e-03 & 1.6702e+1 & 1.79e+05 & 2.8903e+01 & 3.09e+05 \\
 1 & 3.74e-04 & 1.6700e+01 & 4.47e+06 & 2.8900e+01 & 7.73e+06
\end{tabular}
\end{ruledtabular}
\end{table}

\section{Effective decay rate in a stellar plasma}
\begin{figure}[htb]
	\centering
		\includegraphics[width=0.45\textwidth]{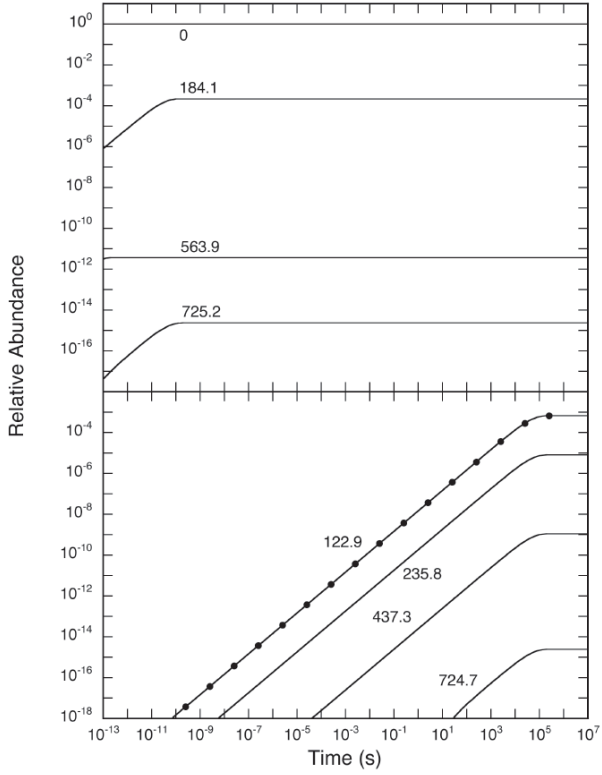}
	\caption{The time evolution of the states shown in Figure~\protect\ref{fig:equilstates} for \tnine\ = 0.25 and $\beta^{2}$ = 9.40 $\times$ 10$^{-3}$\%, assuming no destruction of \Lu\ via $\beta$-decay.The top panel shows states in the ground-state cluster, labeled by their excitation energies (in keV). The isomer cluster is shown in the bottom panel. The two clusters equilibrate after about 10$^{5}$ s, which is ~14-15 orders of magnitude longer than the equilibrium times within the clusters. The dots show the abundance of the isomer calculated using the 2-level approximation described in the text.}
	\label{fig:fullequil}
\end{figure}

The time evolution of the states shown in Figure~\ref{fig:equilstates} can be calculated by solving the set of coupled differential equations that describe the behavior of each state. Since various transition rates can differ by many orders of magnitude at $s$-process temperatures, significant care must be exercised to control the size of numerical errors. This becomes cumbersome once \Lu\ is placed into a larger network of nuclei. However, in a stellar plasma, \Lu\ will rapidly evolve into a configuration of 2 quasi-equilibrium clusters, one of which contains the states in the ground-state and $K$ = 6 bands, while the other is composed of the states in the isomer band. Within the clusters, the states rapidly achieve thermal equilibrium, as described in Sec. II.  To illustrate this point, we have calculated the time evolution of the states shown in Figure~\ref{fig:equilstates} for \tnine\ = 0.25 and mixing amplitude $\beta^{2}$ = 9.40 $\times$ 10$^{-3}$\%. Only the 184.1-keV and 388.9-keV states have tabulated lifetimes~\cite{endsf} and so those for the remaining states were calculated using the procedures described above. The results are shown in Figure~\ref{fig:fullequil}, assuming that only the ground state is populated at $t$ = 0 and that there is no destruction of \Lu\ via $\beta$-decay. The top panel shows the states that equilibrate quickly with the ground state. This cluster reaches thermal equilibrium  within about 0.1 ns. The bottom panel shows the cluster built on the isomer, which reaches equilibrium somewhat later than the ground-state cluster (here the steady-state abundances are reached after about 1 ns) . The rise in the abundances of these states is a consequence of leakage out of the ground-state cluster. At about 10$^{5}$ s, the isomer cluster reaches equilibrium with the ground-state cluster and all of the relative abundances assume their steady-state values. This time scale is about 14-15 orders of magnitude longer than the time scales for equilibration within the clusters. Thus, on the time scale of equilibration between the ground state and the isomer, equilibration within a cluster can be considered to be instantaneous. Therefore, the only $\gamma$-ray transitions that have to be considered explicitly are those between the clusters, i.e. 5$\leftrightarrow$6 and 3$\leftrightarrow$7.

\begin{figure*}[htb]
	\centering
		\includegraphics[width=0.75\textwidth]{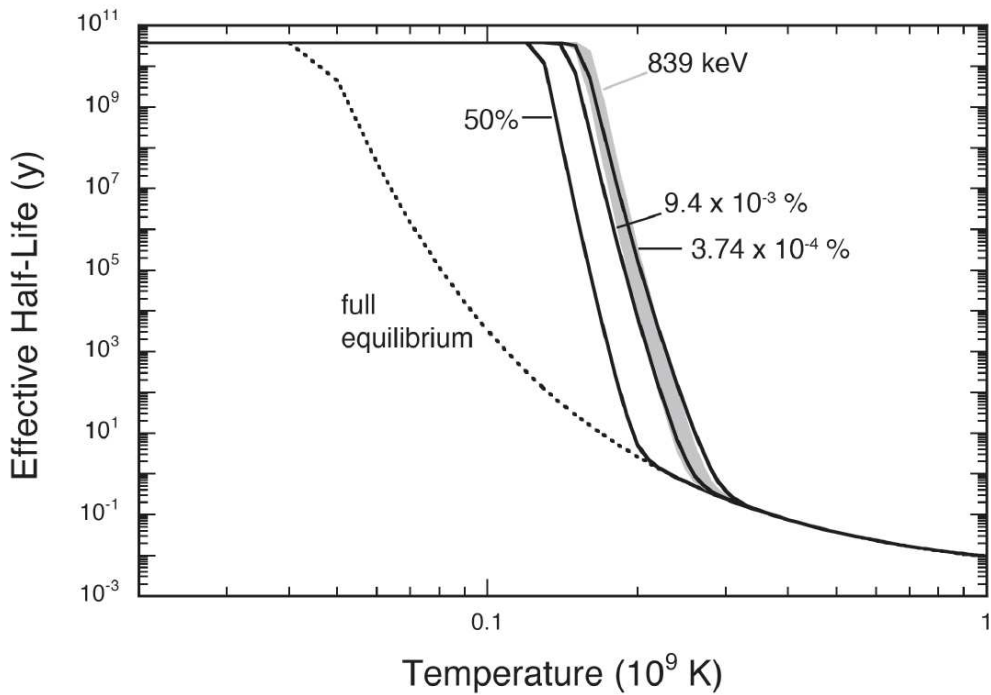}
	\caption{The effective half-life of the ground state of \Lu\ as a function of temperature. The results for $K$-mixing, assuming mixing fractions of 3.74 $\times$ 10$^{-4}$\%, 9.40 $\times$ 10$^{-3}$\%, and  50\%, are represented by the solid lines. The shaded area is the effective half-life if \Lu\ equilibrates via the 838.6-keV state. Full thermal equilibrium is denoted by the dashed line.}
	\label{fig:teff}
\end{figure*}

In this approximation, the time evolution of the clusters can be described by 2 coupled equations:
\begin{eqnarray}
\frac{d}{dt}\left(\sum_{i} n_{i}\right)_{g} = -n_{1}\lambda_{\beta}^{1} - n_{3}\lambda_{37} - n_{5}\lambda_{56} \nonumber \\
+ n_{6}\lambda_{65} +  n_{7}\lambda_{73},
\label{eqn:evol1}
\end{eqnarray}
which describes the ground-state cluster and
\begin{eqnarray}
\frac{d}{dt}\left(\sum_{i} n_{i}\right)_{m} = -n_{2}\lambda_{\beta}^{2} -  n_{6}\lambda_{65} -  n_{7}\lambda_{73} \nonumber \\
 +  n_{3}\lambda_{37} + n_{5}\lambda_{56}
\label{eqn:evol2}
\end{eqnarray}
for the isomer cluster. Here $n_{i}$ refers to the abundance of state $i$ in Figure~\ref{fig:equilstates}; The subscripts {\em g} and {\em m} denote the equilibrium clusters built on the ground state and isomer, respectively; $\lambda_{ij}$ are the $i \rightarrow j$ transition rates, and $\lambda_{\beta}^{1}$ and $\lambda_{\beta}^{2}$ are the laboratory $\beta$-decay rates for the ground state and isomer. Although the $\beta$-decay rates depend on the degree of ionization, the temperatures and densities of the $s$-process are low enough that this effect may be neglected.
The transition rate linking an initial
state \textit{i} with a higher-lying final state \textit{f} via absorption of a photon is
\begin{equation}
    \lambda_{if} = \frac{g_{f}}{g_{i}} \frac{\lambda^{s}_{fi}}{e^{E_{if}/kT}-1}~,
\end{equation}
where $g = 2J+1$, $\lambda^{s}_{fi}$ is the rate for spontaneous decay from
\textit{f} to \textit{i}, $E_{if}$ is the energy difference between
\textit{i} and \textit{f}, and $k$ is Boltzmann's constant. The
corresponding total decay rate between \textit{i} and \textit{f}
(including stimulated emission) is
\begin{equation}
    \lambda_{if} =  \frac{\lambda^{s}_{fi}}{1-e^{-E_{if}/kT}}~.
\end{equation}
Since the states within a cluster are in thermal equilibrium, their relative abundances are given by
\begin{equation}
C_{ij} \equiv \frac{n_{i}}{n_{j}} = \frac{g_{i}}{g_{j}}e^{-(E_{xi}-E_{xj})/kT}.
\label{eqn:ratio}
\end{equation}
Therefore, the abundances appearing in Eqns.~\ref{eqn:evol1} and~\ref{eqn:evol2} can be written as fractions of the ground state or isomer abundances. This allows these equations to be recast as a
concise 2x2 coupled system of equations for the abundances of the ground state and isomer:
\begin{equation}
\frac{d}{dt}n_{1}=a_{11}n_{1}+a_{12}n_{2}
\label{eq:2clustersimple1}
\end{equation}
\begin{equation}
\frac{d}{dt}n_{2}=a_{21}n_{1}+a_{22}n_{2}
\label{eq:2clustersimple2}
\end{equation}
or equivalently as a matrix equation
\begin{equation}
\frac{d}{dt}\mathbf{n} = \mathbf{A}\cdot\mathbf{n}.
\label{eqn:amatrix}
\end{equation}
The transition matrix elements $a_{ij}$ are
\begin{equation}
\begin{array}{clrr}
a_{11}\equiv\frac{\left(-\lambda_{\beta}^{1}-C_{51}\lambda_{56}-C_{31}\lambda_{37}\right)}{\left(1+C_{31}+C_{51}+...\right)} &  
a_{12}\equiv\frac{\left(C_{72}\lambda_{73}+C_{62}\lambda_{65}\right)}{\left(1+C_{31}+C_{51}+...\right)}  \\
a_{21}\equiv\frac{\left(C_{51}\lambda_{56}+C_{31}\lambda_{37}\right)}{1+\left(C_{42}+C_{62}+C_{72}+...\right)} &
a_{22}\equiv\frac{\left(-\lambda_{\beta}^{2}-C_{72}\lambda_{73}-C_{62}\lambda_{65}\right)}{1+\left(C_{42}+C_{62}+C_{72}+...\right)},
\end{array}
\end{equation}
where $a_{12}$ and $a_{21}$ represent the effective $m \rightarrow g$ and $g \rightarrow m$ rates, respectively.
The denominators in these expressions are formally a sum over all of the $C_{ij}$ for each cluster. In practice, only the lowest-energy states contribute to this sum and for \sproc\ temperatures, $C_{ij} \ll$ 1.
Eq.~\ref{eqn:amatrix} has the solution
\begin{equation}
\mathbf{n}(t) = \mathbf{N}_{1} e^{-\lambda_{1}t} + \mathbf{N}_{2}
e^{-\lambda_{2}t}
\label{eqn:abund}
\end{equation}
where \textbf{N}$_{i}$ and --$\lambda_{i}$ are the eigenvectors 
(normalized to the correct initial abundances) and eigenvalues
of \textbf{A}, respectively. The smallest eigenvalue corresponds to the effective decay rate of the (quasi) equilibrated nucleus. The other eigenvalue (which is approximately equal to $a_{22}$) is the inverse of the time constant for quasi or full equilibration. The resulting effective half-life of \Lu\ is shown in Figure~\ref{fig:teff}. We have also calculated the effective half-life for equilibration through the 838.6-keV state, using $\tau$ = 10--300 ps for this state~\cite{doll}, which is consistent with what is reported by Klay {\em et al.}~\cite{klay91b}.
As shown in Figure~\ref{fig:fullequil}, this two-level approximation accurately accounts for the evolution of the isomer.

\section{Astrophysical aspects}
Assuming that the $K$-mixing scenario is in fact valid, our calculation shows that the onset of quasi-equilibrium and the transition to full equilibrium may occur at lower temperatures than previously thought~\citep{klay91b}. In particular, \Lu\ will reach full thermal equilibrium for \tnine\ $\gtrsim$ 0.22-0.30  for the range of mixing considered here. In contrast full equilibrium through the 838.6-keV state occurs for \tnine\ $\approx$ 0.29-0.32. This is not too surprising since we assume that the mediating level is at a lower excitation energy: 725 keV compared to 839 keV. However, it is interesting to note that a mixing amplitude of only ~9.40 $\times$ 10$^{-3}$\% is sufficient to compete with equilibration via the 838.6-keV state. Since the corresponding mixing matrix element of 5 eV is on the order of atomic interactions, it is fair to wonder if any nuclear state can be considered pure. In fact, the recent observation of $\Delta K$ = 13 mixing in $^{174}{\rm Lu}$ with V $\approx$ 19 eV suggests a minimum V in the range of 10-20 eV~\cite{drac}. This general conclusion likely applies to other $K$-isomers in the \sproc\ (e.g. $^{180}{\rm Ta}$ \cite{wal}). In other words, the degree of $K$-mixing that exists in rotational nuclei is probably sufficient to allow $K$-isomers to reach at least quasi-equilibria with their ground states during some phases of the \sproc. In the case of \Lu, quasi- or full equilibrium is reached for temperatures characteristic of the thermal-pulse phase in low-mass AGB stars \cite{bus} or the weak \sproc\ in massive stars \cite{pet}. In both cases, the major neutron source is the $^{22}{\rm Ne}$($\alpha$,n)$^{25}{\rm Mg}$ reaction.

\begin{figure}[htb]
	\centering
		\includegraphics[width=0.45\textwidth]{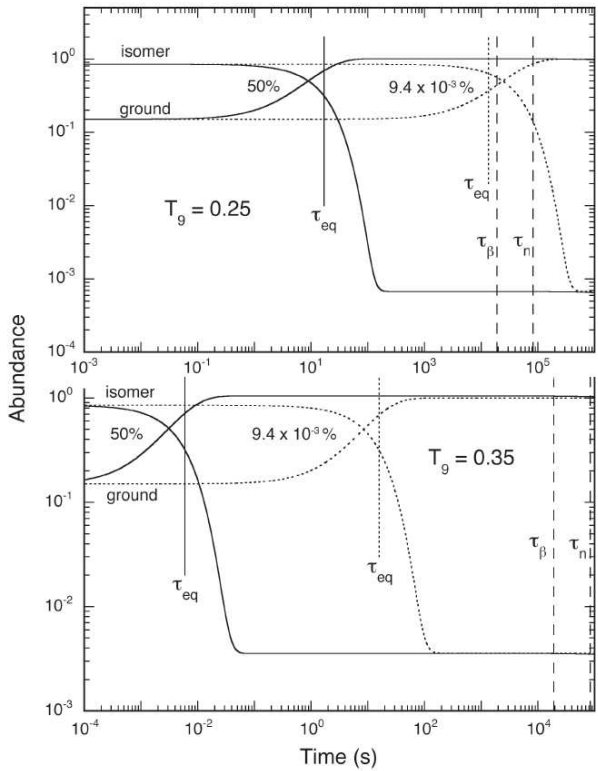}
	\caption{The time evolution of the isomer and ground state - assuming no production or destruction - for \tnine\ = 0.25 and 0.35 and $K$-mixing amplitudes of 9.40 $\times$ 10$^{-3}$\% (dotted lines) and 50\% (solid lines). The time scales $\tau_{eq}$, $\tau_{n}$ and $\tau_{\beta}$ are denoted by vertical lines. For \tnine\ = 0.35, $\tau_{eq}$ is much shorter than either $\tau_{n}$ or $\tau_{\beta}$, which indicates that the equilibrating transitions are fast enough to maintain thermal equilibrium in the presence of n-capture and $\beta$-decay. However, this is not the case for 9.40 $\times$ 10$^{-3}$\% mixing at \tnine\ = 0.25. }
	\label{fig:time}
\end{figure}

The effective half-life shown in Figure \ref{fig:teff} gives an indication of the temperature range where equilibration will occur, but the actual behavior of \Lu\ in a stellar environment will depend on the competition between the time scale for equilibration, $\tau_{eq}$ (the inverse of the largest eigenvalue in Eqn.~\ref{eqn:abund}) and those for neutron capture, $\tau_{n}$ and $\beta$-decay, $\tau_{\beta}$. If $\tau_{eq}$ is smaller than both $\tau_{n}$ and $\tau_{\beta}$, then \Lu\ will maintain thermal equilibrium with an effective half-life as shown in Figure \ref{fig:teff} . Otherwise, \Lu\ will achieve a quasi-equilibrium in which the relative population of excited states can be quite different from the equilibrium values. In this case, the approach to equilibrium will lag behind abundance changes from n-capture and $\beta$-decay and it is necessary to include the equilibrating transitions ($a_{12}$ and $a_{21}$ from Eqns.~\ref{eq:2clustersimple1} and~\ref{eq:2clustersimple1}) explicitly when calculating the time evolution of \Lu. During thermal pulses in low-mass AGB stars (with peak temperatures of \tnine\ = 0.25-0.35 and neutron densities of 8 $\times$ 10$^{6}$ -- 3 $\times$ 10$^{10}$ cm$^{-3}$), the question of whether full or quasi-equilibrium pertains depends strongly on the amplitude for $K$-mixing. This is shown in Figure~\ref{fig:time}. For \tnine = 0.35, \Lu\ will be in full thermal equilibrium for the entire range of $K$-mixing considered here because $\tau_{eq}$ $\ll$ $\tau_{n}$ and $\tau_{\beta}$ (here we have assumed a neutron density of 3 $\times$ 10$^{10}$ cm$^{-3}$). However, this is not the case in the limit of weak $K$-mixing and lower temperatures. At \tnine\ = 0.25, $\tau_{eq}$ $\approx$ $\tau_{\beta}$ for $\beta^{2}$ = 9.40 $\times$ 10$^{-3}$\% and a quasi-equilibrium will ensue. A similar situation pertains for equilibration through the 838.6-keV state. For T $\lesssim$ 0.2, $\tau_{eq}$ $>$ $\tau_{\beta}$ even in the case of maximal mixing and thus \Lu\ can not be ascribed an effective decay rate.

\begin{figure*}[htb]
	\centering
		\includegraphics[width=0.8\textwidth]{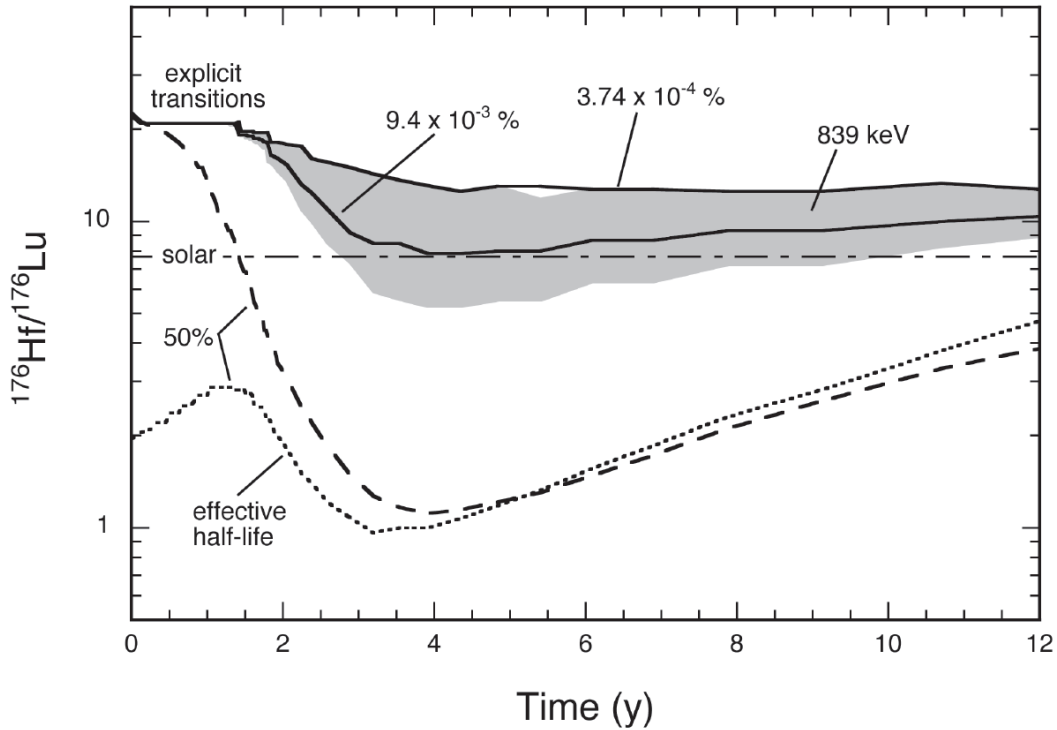}
	\caption{The evolution of \Hf/\Lu\ as a function of time since the start of the neutron burst for the 9$^{th}$ thermal pulse of a 3 M$_{\odot}$ AGB star of solar composition. The solar ratio at the time of formation of the solar system is indicated by the horizontal dashed line. The ratios for 3.74 $\times$ 10$^{-4}$\%, 9.40 $\times$ 10$^{-3}$\% and 50\% $K$-mixing, calculated by including the transitions linking the ground state and isomer (labeled as explicit transitions) are represented  by the solid and heavy dashed lines, respectively. Ratios calculated using the effective half-lives from Fig.~\protect\ref{fig:teff} deviate significantly these results and are shown as the short-dased and dotted lines . The grey shaded area is the \Hf/\Lu\ ratio for equilibration through the 838.6-keV state for lifetimes of 10 ps (lower bound) and 300 ps (upper bound), which was also calculated using the equilibrating transitions.}
	\label{fig:network}
\end{figure*}

To show schematically how the abundance \Lu\ will evolve during the \sproc\, we have performed post-processing calculations for a 3 M$_{\odot}$ AGB star of solar composition during the thermal-pulse phase. The temperature and density profiles at the base of the convective shell as well as the abundances of the major isotopes were calculated using the Monash version of the Mount
Stromlo stellar-structure  code (Ref.~\citep{latt} and references therein). The initial abundances of the \sproc\ isotopes were calculated for temperatures and densities appropriate to the interpulse phase, during which the  $^{13}{\rm C}$($\alpha$,n)$^{16}{\rm O}$ reaction is the primary source of neutrons. Neutron-capture rates in the Yb-Hf region were taken from the {\sc kadonis} compilation~\cite{kad}. Since both \Hf\ and \Lu\ are produced almost entirely by the \sproc\ (there is a small $p$-process contribution to \Hf, which we will ignore), the ratio \Hf/\Lu\ is a useful \sproc\ diagnostic because its value at the time of the formation of the solar system is reasonably well established. We have adopted the value 7.67(18), which is obtained from the tabulated abundances of \Lu\ and \Hf~\cite{anders89}, corrected for the radioactive decay of \Lu\ (with $T_{1/2}$ = 37.6 Gy) over the past 4.55 $\times$ 10$^{9}$ years. The results of our calculation for the 9$^{th}$ thermal pulse are shown in Figure~\ref{fig:network}. Here, the peak temperature is \tnine\ $\simeq$ 0.24 and thus the $^{22}{\rm Ne}$ neutron source is only marginally activated. The neutron burst lasts for about 12 years, which is about 10\% of the duration of the entire convective phase. These results are not meant to represent a quantitative depiction of the \sproc\ for this star. We have calculated the \sproc\ for conditions found only at the base of the convective shell and have not included mixing. Our purpose is simply to show how \Hf/\Lu\  behaves for different degrees of equilibration (a more realistic treatment of nucleosynthesis can be found in Ref.~\cite{heil}).
The effective decay rate does a reasonable job of describing the evolution of \Lu\ for the case of maximal mixing (50\%), but only after about 4 y beyond the start of the burst. For the (relatively) low temperatures in this pulse the use of an effective decay rate does not accurately describe the behavior of \Lu\ for lower mixing fractions and thus, the transitions linking the ground-state and isomer clusters must be used instead.  One thing that is clear in Figure~\ref{fig:network} is that \Hf/\Lu\ decreases as mixing becomes more efficient. The same is true if the peak temperature is increased. The reason for this is that the equilibration time drops as the mixing fraction or temperature is increased, which favors production of the ground state of \Lu. High levels of $K$-mixing appear to be ruled out because they lead to too much \Lu\ relative to \Hf. As pointed out above, minimal amounts of $K$-mixing within the scenario outlined here are sufficient to compete with the observed transitions via the 838.6-keV state. The actual amount of mixing required depends on temperature, but a mixing fraction on the order of 3.74 $\times$ 10$^{-4}$\% -- 0.015\% or equivalently, a mixing matrix element of only 1--6.6 eV is sufficient throughout the range of temperatures for the $^{22}{\rm Ne}$ neutron source. As shown in Table~\ref{tab:results}, the branching ratios for the relevant $\gamma$-ray transitions are probably too small to be observed, which complicates the use of \Lu\ as an \sproc\ thermometer. The small value required for the matrix element implies that $K$-mixing should at least contribute to the equilibration of \Lu. However, without experimental verification, it may be difficult to predict the amount of \Lu\ produced, or \Hf/\Lu\ for a given temperature regime. On the other hand, if small amounts of $K$-mixing can have a dramatic impact on equilibration, then it seems plausible that this mechanism will operate in other places in the \sproc.

\begin{acknowledgments}
This work was supported in part by USDOE grants \#DE-FG02-97ER4104 and \#DE-AC02-06CH11357. We would like to thank J. Engel for advice and useful discussions. One of the authors (AEC) wishes to thank Argonne National Laboratory for their hospitality and another (RL) would like to thank J. Lattanzio for his assistance with the Monash stellar-structure code.
\end{acknowledgments}

\bibliographystyle{prsty}
\bibliography{176Lu-1}
\end{document}